\documentstyle[aps,pre,eqsecnum]{revtex}

\draft

\begin{document}
\title{Instantons and Intermittency}

\author{G. Falkovich$^a$, I. Kolokolov$^b$, V. Lebedev$^{a,c}$ and
A.Migdal$^d$}
\address{$^a$ Physics Department, Weizmann Institute of Science,
Rehovot 76100, Israel \\ $^b$ Budker Institute of
Nuclear Physics, Novosibirsk 630090, Russia
\\ $^{c}$ Landau Inst. for Theor. Physics, Moscow, Kosygina 2,
117940, Russia\\$^d$ Physics Department, Princeton University,
Princeton NJ 08544-1000 USA}
\maketitle

\begin{abstract}
We propose the new method for finding the non-Gaussian tails of
probability distribution function (PDF) for solutions of a stochastic
differential equation, such as convection equation for a passive scalar,
random driven Navier-Stokes equation etc. Existence of such
tails is generally regarded as a manifestation of intermittency
phenomenon. Our formalism is based on the WKB approximation in the
functional integral for the conditional probability of large fluctuation.
We argue that the main contribution to the functional integral
is given by a coupled field-force configuration -- {\em instanton}.
As an example, we examine the correlation functions of the passive scalar
$u$ advected by a large-scale velocity field $\delta$-correlated
in time. We find the instanton determining the tails of the generating
functional and show that it is different from the instanton that determines
the probability distribution function of high powers of $u$. We discuss the
simplest instantons for the Navier-Stokes equation.
\end{abstract}

\section{Introduction}

The intermittency phenomenon (reflected in non-Gaussian, scaling-violating
tails of PDF) is believed to be the hardest part of the yet to be built theory
of turbulence. It is related to rare fluctuations which cannot be treated
in terms of a perturbation theory. Neither the physical mechanism nor the
mathematical properties of such fluctuations are known.

Now, what is the most likely force which can lead to the given rare
fluctuation of the field? The main idea of this paper is that such
force is not random at all. It satisfies the well defined equation,
which follows from the WKB approximation in the functional integral.
Asymptotically, the fluctuations of the force around this most likely
one are negligible. In this respect, the method is similar to
the ``optimal fluctuation'' method used at treating properties
of a solid with quenched disorder (see e.g. the book \cite{ES}) or
at treating high-order terms of the perturbation series in the quantum
field theory \cite{Lip}.

The problem under consideration is quite general, it can be formulated
for any field governed by a nonlinear dynamic equation and driven by a
random ``force''. Generally, the PDF
of the field depends both on the statistics of the driven force and
on the form of the dynamical equation. Here, we are interested in the second
dependence so that we assume the force to be Gaussian. Because of
nonlinearity, the PDF of the field is non-Gaussian even for a Gaussian
random force. Note that strong intermittency appears also for linear
problems with ``multiplicative noise'', for instance, for a passive scalar
advected by a random velocity field.

We start with the dynamical equation
\begin{equation}
\partial_tu+{\cal L}(u)=\phi \,,
\label{i1}\end{equation}
that controls the evolution of a field $u(t,{\bf r})$ under the
action of a random ``force'' $\phi(t,{\bf r})$. Here ${\cal L}(u)$
is a nonlinear expression, it can be thought to be local in space.
Generally, both the field $u$ and the force $\phi$ have a number of
components. The Gaussian  statistics of the force $\phi$ is
completely characterized by the pair correlation function
\begin{equation}
\langle\phi(t_1,{\bf r}_1)\phi(t_2,{\bf r}_2)\rangle=
\Xi(t_1-t_2, {\bf r}_1-{\bf r}_2) \,.
\label{i2} \end{equation}
In principle, the relations (\ref{i1},\ref{i2}) contain all the information
about the statistics of $u$.

The equation (\ref{i1}) describes e.g. thermal fluctuations in hydrodynamics
where it is reduced to the well known Langevin equation \cite{LL}.
Then $\phi$ is short-correlated in time and in space that is
it can be treated as a white noise. For some systems, this thermal
noise produces remarkable dynamical effects. Some examples are collected
in the book \cite{93KL}. Here, we will be interested in turbulence where
$\phi$ is an external ``force'' correlated on large scales in space.
Turbulence was first treated in terms of the equation
(\ref{i1}) by Wyld \cite{61Wyl} who formulated the diagram technique
as a perturbation series with respect to the nonlinear term in the
Navier-Stokes equation.
The diagram technique cannot be applied to our problem since we are
interested in non-perturbative effects. Nevertheless we can use the
functional that generates the technique since it is a non-perturbative
object. Such generating functional was introduced in \cite{76Dom,76Jan},
for the equation (\ref{i1}) it has the form
\begin{eqnarray} &&
{\cal Z}(\lambda)\equiv
\left\langle
\exp\left(i\int dt\, d{\bf r}\, \lambda u \right)
\right\rangle
\nonumber \\ &&
=\int {\cal D}u{\cal D}p
\exp\left(i{\cal I} +i\int dt\, d{\bf r}\, \lambda u \right) \,,
\label{i3} \end{eqnarray}
where $p$ is an auxiliary field and the effective action is
\begin{eqnarray} &&
{\cal I}=\int dt\, d{\bf r}\, p\bigl[\partial_t u+{\cal L}(u)\bigr]
\nonumber \\ &&
+{i\over2}\int dt\, dt'\, d{\bf r}\, d{\bf r}'\,
\Xi(t-t',{\bf r}-{\bf r}') p p' \,.
\label{i4} \end{eqnarray}
The coefficients of the expansion of ${\cal Z}$ in $\lambda$ are the
correlation functions of $u$. The auxiliary field $p$ determines
response functions of the system, for instance, the linear response
function (Green function) is $G=\langle u p\rangle$. Note the
remarkable property \cite{DP78}
$$ \int {\cal D}u{\cal D}p\exp\left(i{\cal I}\right)=1 \ ,$$
related to causality. That is the reason why the normalization constant
is unity in (\ref{i3}). This makes it possible to average directly ${\cal Z}$
over any additional random field if necessary.

The asymptotics of ${\cal Z}(\lambda)$ at large $\lambda$ is determined
by the saddle-point configuration (usually called classical trajectory
or instanton) which should satisfy the following equations obtained
by varying the argument of the exponent in (\ref{i3}) with
respect to $u$ and $p$
\begin{eqnarray} &&
\partial_tu+{\cal L}(u)=
-i\int dt'\, d{\bf r}'\, \Xi(t-t',{\bf r}-{\bf r}')p(t',{\bf r}') \,,
\label{i5} \\ &&
\partial_t p-{\delta{\cal L}\over\delta u}p=\lambda \,.
\label{i6} \end{eqnarray}
Solutions of the equations (\ref{i5},\ref{i6}) are generally smooth
functions of $t$ and ${\bf r}$. Comparing the equations (\ref{i1})
and (\ref{i5}) we conclude that the right hand side of the equation
(\ref{i5}) just describes a special force configuration necessary
to produce the instanton. If $u_{\rm inst}$ is a solution of
(\ref{i5},\ref{i6})
then asymptotically at large $\lambda$
\begin{equation}
\delta\ln{\cal Z}(\lambda)/\delta\lambda=
iu_{\rm inst} \ .
\label{ij} \end{equation}

Let us discuss the boundary conditions for the saddle-point equations.
The equation (\ref{i5})
implies that we should fix the value $u_{\rm in}$ for the field
$u$ at the initial time $t_{\rm in}$. Contrary, a boundary condition for the
field
$p$ is implied at the remote future since, as follows from (\ref{i6}), it
propagates in the negative direction in time. Minimization of the action
generally requires $p\rightarrow0$ at $t\rightarrow\infty$. For the
instantons discussed below, the finiteness of the action will also require
$u\rightarrow0$ at $t\rightarrow-\infty$.

If one is interested in the simultaneous statistics of $u$ then the
function $\lambda$ can be chosen as
\begin{equation}
\lambda(t,{\bf r})=y\delta(t)\lambda_0({\bf r}) \,,
\label{b4} \end{equation}
where $y$ is a number and $\lambda_0$ is an appropriate function of
${\bf r}$ depending on what spatial correlation functions we are
going to study. In this case, we should find the solution for $p$
satisfying the rule: $p=0$ at $t>0$.
The system (\ref{i5},\ref{i6}) is thus to be treated  for $t<0$ only.
That corresponds to the causality principle since only processes
occurring in the past could influence the value of the simultaneous
correlation functions at $t=0$. The formal ground for the rule
follows from the consideration of the problem in the restricted time
interval $t<t_0$ what is possible if $\lambda=0$ at $t>t_0$. Then the
minimization of ${\cal I}+\int dt\, d{\bf r}\,\lambda u$ over the final
value $u(t_0)$ gives $p(t_0)=0$ because of the boundary term originating
from $\int dt\, d{\bf r}\, p\partial_t u$.

One may be interested also in the probability distribution function
${\cal P}(u)$ for the field $u$. It can be expressed via the generating
functional ${\cal Z}(\lambda)$ via the functional Fourier transform
\begin{equation}
{\cal P}(u)=\int {\cal D}\lambda\,{\cal Z}(\lambda)
\exp\left(-i\int dt\, d{\bf r} \, \lambda u \right) \,.
\label{i7} \end{equation}
We expect that the behavior of ${\cal P}(u)$ at large $u$ as well as
the behavior of ${\cal Z}(\lambda)$ for large $\lambda$ is associated
with some saddle-point configurations. Generally, the configurations
are not always the same for both (\ref{i3}) and (\ref{i7}).
Indeed, we see from (\ref{i7}) that the tail of ${\cal P}(u)$
at large $u$ corresponds to a large value
of $\delta \ln {\cal Z}(\lambda)/\delta\lambda$ which is related
to large $\lambda$ only if the tails of both PDF and the generating
functional decay faster  than exponent -- see the example in Sect.
\ref{sec:nse}.
Otherwise, those tails are determined by different configurations as
is demonstrated in Section \ref{sec:pas}.

The best starting point to develop the instanton formalism is the problem
of white-advected passive scalar $\theta$ since it allows for a
detailed analytical treatment \cite{74Kra-a,94SS,95CFKLa}. It will
be shown in Section \ref{sec:pas}
that both ${\cal P}(\theta)$ and ${\cal Z}(\lambda)$ have
exponential tails as it has been established before by Shraiman and Siggia
\cite{94SS} (see also \cite{95CFKLa}). By using this example, we
shall explicitly demonstrate that different instanton configurations are
responsible for the tails of the generating functional ${\cal Z}(\lambda)$
at large $\lambda$ and of PDF at large $\theta$ respectively. It is
instructive to recognize the difference between the instantons:
We shall show that the instanton that is responsible for large $\theta$
corresponds to a small strain and suppressed stretching. Contrary,
the instanton that determines the tails of ${\cal Z}$ corresponds
to a large value of strain.

Section \ref{sec:nse} presents the first step in studying instantons of
the Navier-Stokes equation. Only instantons for the two-point generating
functional $\langle\exp(i\lambda(u_1-u_2)\rangle$ will be considered.
The family of such instantons corresponds to the velocity fields with
a linear spatial profile at $r\ll L$. Consideration of the
instanton perturbations (giving the fluctuation contribution into the action)
that correspond to spiral creation in the straining field of the instanton will
be the subject of further publications.

\section{Passive Scalar Advected by a Large-scale Velocity Field}
\label{sec:pas}

Let us show how the general formalism described in the Introduction
works for a particular problem: the advection of a passive scalar field
$\theta(t,{\bf r})$ by an incompressible turbulent flow
in $d$-dimensional space \cite{68Kra-a,74Kra-a,94SS,95CFKLa}.
The advection is governed by the following equation
\begin{equation}
(\partial_t +v_{\alpha}\nabla_{\alpha}-\kappa \triangle)
\theta=\phi\,,\quad \nabla_\alpha v_\alpha=0\,,
\label{a1} \end{equation}
where $\phi(t,{\bf r})$ is the external source, {\bf v} is the
advecting velocity and $\triangle$ designates Laplacian,
$\kappa$ being the diffusion coefficient. Both ${\bf v}(t,{\bf r})$
and $\phi(t,{\bf r})$ are random functions of $t$ and ${\bf r}$.
We regard the statistics of the velocity and of the source to be independent.
Therefore, all correlation functions of $\theta$ are to be treated as
averages over both statistics.

We assume that the source $\phi$ is $\delta$-correlated
in time and spatially correlated on a scale $L$ and has a Gaussian statistics
completely determined by the pair correlation function:
\begin{equation}
\langle{\phi(t_1,{\bf r}_1)\phi(t_2,{\bf r}_2)}\rangle=
\delta(t_1-t_2)\chi(r_{12})\ .
\label{a2} \end{equation}
Here $\chi(r_{12})$ as a function of the argument
$r_{12}\equiv|{\bf r}_1-{\bf r}_2|$
decays on the scale $L$. We are interested in the behavior of the
correlation functions on scales $r\ll L$. Thus, only the constant
$P_2=\chi(0)$ will enter all the answers. The constant $P_2$ has the
physical meaning of the production rate of $\theta^2$.

Following Kraichnan \cite{68Kra-a,74Kra-a}, we consider the case of a
Gaussian velocity ${\bf v}$ delta-correlated in time and containing only
large-scale space harmonics. Then the velocity statistics is
also completely determined by the pair correlation function
\begin{eqnarray} &&
\langle  v_\alpha(t_1,{\bf r}_1) v_\beta(t_2,{\bf r}_2)\rangle=
\delta(t_1-t_2)V_{\alpha\beta}\,,
\nonumber \\ &&
V_{\alpha\beta}=V_0
\delta_{\alpha\beta}-{\cal K}_{\alpha\beta}({\bf r}_{12}),\quad
{\cal K}_{\alpha\beta}(0)=0\ .
\label{a5} \end{eqnarray}
Here the so-called eddy diffusivity is as follows
\begin{equation}
{\cal K}_{\alpha\beta}={D}
(r^2\delta_{\alpha\beta}-r_\alpha r_\beta)+\frac{D(d-1)}{2}
\delta_{\alpha\beta} r^{2}\,,
\label{a6} \end{equation}
where $d$ is the dimensionality of space and isotropy of the velocity
statistics being assumed. The representation (\ref{a5},\ref{a6}) is valid
for the scales less than the velocity infrared cut-off $L_u$, which is
supposed to be the largest scale of the problem. Then $V_0$ and
${\cal K}_{\alpha\beta}$ in (\ref{a5}) are two first terms of
the expansion of the velocity correlation function in $r/L_u$ so that
$D\sim V_0/L_u^2$. We presume also the inequality $dDL^2\gg\kappa$
which guarantees the existence of a convective interval of scales
$r_d\ll r\ll L$ where correlation functions of the passive scalar
are formed mainly by stretching in the velocity field. Here
$r_d=2\sqrt{\kappa/(D(d-1))}$ is the diffusion length.

The statistics of the large-scale velocity field has the remarkable property:
It follows from the expressions (\ref{a5},\ref{a6}) that the correlation
function of the strain field $\sigma_{\alpha\beta}=\nabla_\beta v_\alpha$
is ${\bf r}$-independent:
\begin{equation}
\langle\sigma_{\alpha\beta}(t_1)
\sigma_{\mu\nu}(t_2)\rangle=
D\left[(d+1)\delta_{\alpha\mu}\delta_{\beta\nu}
-\delta_{\alpha\nu}\delta_{\beta\mu}
-\delta_{\alpha\beta}\delta_{\mu\nu}\right]
\delta(t_1-t_2)  \,,
\label{ax6} \end{equation}
That means that the strain field $\sigma_{\alpha\beta}$ can be treated as a
random function of time $t$ only. Just that property enables one to find
in detail statistical properties of the field $\theta$ \cite{94SS,95CFKLa}.
To exploit the property, it is convenient to pass into the comoving
reference frame that is to the frame moving with the velocity of a
Lagrangian particle of the fluid. That means that we pass to the new
space variable ${\bf r}-{\bbox\varrho}(t)$ where ${\bbox\varrho}(t)$
is the Lagrangian trajectory of the particle \cite{87BL,91Lvo}. We
will take the particle positioned at the origin at time $t=0$, then
\begin{equation}
{\bbox\varrho}(t)=\int\limits_{0}^{t}
d\tau \, {\bf v}\big(\tau,\bbox{\varrho}(\tau)\big) \,.
\label{traj} \end{equation}
After the transformation ${\bf r}\to{\bf r}-{\bbox\varrho}(t)$, the equation
(\ref{a1}) acquires the form
\begin{equation}
\left\{\partial_t +
\bigl[v_{\alpha}(t,{\bf r})-v_{\alpha}(t,0)\bigr]
\nabla_{\alpha}-\kappa \triangle\right\}\theta=\phi\,,
\label{an1} \end{equation}

It is seen from (\ref{a5},\ref{a6}) that the statistics of
$v_{\alpha}(t,{\bf r})-v_{\alpha}(t,0)$ coincides with the
statistics of $\sigma_{\alpha\beta}r_\beta$. That means that the
generating functional corresponding to (\ref{an1}) can be written as
\begin{equation}
{\cal Z}(\lambda)=\int{\cal D}\theta\,{\cal D}p\,{\cal D}\sigma\,
\exp\left(-{\cal F}(\sigma)+i{\cal I}+i\int dt\,d{\bf r}\,\lambda\theta\right)
\,.
\label{an2} \end{equation}
where $\sigma_{\alpha\beta}$ is a function of time satisfying
$\sigma_{\alpha\alpha}=0$, its PDF is $\exp(-{\cal F})$.
The effective action $I$ and the functional ${\cal F}$ in (\ref{an2}) are
\begin{eqnarray} &&
i{\cal I}=i\int dt\,d{\bf r}\,(p\partial_t\theta
+p\sigma_{\alpha\beta}r_\beta\nabla_\alpha\theta
+\kappa\nabla p\nabla\theta)
\nonumber \\ &&
-\frac{1}{2}\int dt\,d{\bf r}_1\,d{\bf r}_2\,p_1\chi(r_{12})p_2 \,,
\label{l4} \\ &&
{\cal F}=\frac{1}{2d(d+2)D}\int dt\,
\left[(d+1)\sigma_{\alpha\beta}\sigma_{\alpha\beta}
+\sigma_{\alpha\beta}\sigma_{\beta\alpha}\right] \,.
\label{lx4} \end{eqnarray}
Note that there is the difference between (\ref{i3}) and
(\ref{an2}) which is in the presence of an additional random
field $\sigma_{\alpha\beta}$.

\subsection{Uniaxial Instanton}

Here we examine the saddle-point contribution to the generating functional
${\cal Z}(\lambda)$. The equations describing the saddle points are extremum
conditions for $i{\cal I}+i\int dt\,d{\bf r}\,\lambda\theta-{\cal F}$.
Starting from the expressions (\ref{l4},\ref{lx4}) we find
\begin{eqnarray} &&
\partial_t\theta +
\sigma_{\alpha\beta}r_\beta\nabla_\alpha\theta
-\kappa \nabla^2\theta = -i\int d{\bf r}^\prime\,
\chi(|{\bf r}-{\bf r}^\prime|)p(t,{\bf r}^\prime) \,,
\label{b1} \\ &&
\partial_t p +
\sigma_{\alpha\beta}r_\beta \nabla_\alpha p
+\kappa \nabla^2 p=\lambda \,,
\label{b2} \\ &&
\sigma_{\alpha\beta}(t)=iD\int d{\bf r} \,
\Big((d+1) r_\beta \nabla_\alpha \theta
-r_\alpha\nabla_\beta \theta
-\delta_{\alpha\beta}r_\gamma
\nabla_\gamma\theta\Big)p \,,
\label{b3} \end{eqnarray}
where $p=p(t,{\bf r})$, $\theta=\theta(t,{\bf r})$.
If to take into account only the saddle-point
contribution described by (\ref{b1},\ref{b2},\ref{b3}) then
\begin{equation}
{\cal Z}(\lambda)=
\left\langle\exp\left(i\int dt\,d{\bf r}\,
\lambda\theta\right)\right\rangle\propto
\exp(-{\cal F}_{\rm extr}) \,,
\label{bi4} \end{equation}
where ${\cal F}_{\rm extr}$ is the saddle-point value of
${\cal F}-i{\cal I}-i\int dt\,d{\bf r}\,\lambda\theta$. One get from
(\ref{l4},\ref{lx4},\ref{b2})
\begin{eqnarray} &&
{\cal F}_{\rm extr}
=\frac{1}{2}\int dt\,d{\bf r}_1\,d{\bf r}_2\,p_1\chi(r_{12})p_2
\nonumber \\ &&
+\frac{1}{2d(d+2)D}\int dt\,
\left[(d+1)\sigma_{\alpha\beta}\sigma_{\alpha\beta}
+\sigma_{\alpha\beta}\sigma_{\beta\alpha}\right] \,.
\label{bi3} \end{eqnarray}

In the following, we will be interested in simultaneous correlation
functions of $\theta$ so that
we take the field $\lambda$ in the form (\ref{b4})
and solve the equations for negative time $t<0$. Let us stress that
for the function (\ref{b4}) the term $\lambda\theta$ is not influenced
by the transformation ${\bf r}\to{\bf r}-{\bbox\varrho}(t)$ because of
${\bbox\varrho}(0)=0$. Note that the system of equations
(\ref{b1},\ref{b2},\ref{b3}) with the function (\ref{b4}) is invariant
under the transformation
\begin{equation}
\sigma\to X\sigma\,,\
p\to Xp\,,\ t\to X^{-1}t\,,\
y\to Xy\,,\ \kappa\to X\kappa \,,\
{\cal F}_{\rm extr}\to X{\cal F}_{\rm extr}\,,
\label{bi7} \end{equation}
where $X$ is an arbitrary factor. It leads to the conclusion that
\begin{equation}
{\cal F}_{\rm extr}=y f(y/\kappa) \,,
\label{bi8} \end{equation}
with the function $f$ to be determined.

We will treat nearly single-point statistics.
That means that the space support of the function $\lambda_0$
in (\ref{b4}) is taken to be much smaller than the pumping length $L$. From
the other hand, we would like to avoid bulky formulas related to
the account of diffusion.
Therefore, the size of the support is believed to be much
larger than the diffusion length $r_d$. We thus come to
\begin{equation}
\lambda(t,{\bf r})=y\delta(t)\delta_\Lambda({\bf r}) \,,
\label{bb4} \end{equation}
where $\delta_\Lambda({\bf r})$ is a function with the characteristic
size $\Lambda^{-1}$ satisfying $L\gg\Lambda^{-1}\gg r_d$ and normalized:
$\int d{\bf r}\,\delta_\Lambda({\bf r})=1$. For example, we can take
\begin{equation}
\delta_\Lambda({\bf r})=
\frac{\Lambda^d}{\pi^{d/2}}
\exp(-\Lambda^2r^2) \,.
\label{bb5} \end{equation}
We thus examine the following object
\begin{equation}
{\cal Z}_\Lambda=
\langle\exp(iy\theta_\Lambda)\rangle \,,
\label{bb6} \end{equation}
where
\begin{equation}
\theta_\Lambda=
\int d{\bf r}\, \delta_\Lambda({\bf r})
\theta(t=0,{\bf r}) \,.
\label{bb7} \end{equation}
Having in mind the inequality $\Lambda^{-1}\gg r_d$, we omit in the
following the diffusive terms in the equations.
The extremum conditions (\ref{b1},\ref{b2}) are then as follows:
\begin{eqnarray} &&
\partial_t\theta+\sigma_{\alpha\beta}r_\beta\nabla_\alpha\theta
=-i\int d{\bf r}'\,\chi p' \,,
\label{l5} \\ &&
\partial_t p+\sigma_{\alpha\beta}r_\beta\nabla_\alpha p=
y\delta(t)\delta_\Lambda({\bf r}) \,,
\label{l6}  \end{eqnarray}

It is natural to seek a solution of (\ref{l5},\ref{l6},\ref{b3}) in
the uniaxial form what means that $\sigma_{\alpha\beta}$ is a
diagonal matrix with the components
\begin{equation}
{\rm diag}\,\sigma=(-s, s/(d-1), \dots) \,.
\label{lf1} \end{equation}
As it was suggested in \cite{95CFKLa}, it is useful to pass to the new fields
\begin{equation}
\tilde\theta(t,{\bf r})
=\theta(t,e_\parallel x,e_\perp{\bf r}_\perp) \,, \quad
\tilde p(t,{\bf r})=p(t,e_\parallel x,e_\perp{\bf r}_\perp) \,,
\label{lf2} \end{equation}
where $x$ is the coordinate along the marked direction, ${\bf r}_\perp$
is the component of the radius-vector ${\bf r}$ perpendicular to the
direction and
\begin{equation}
e_\parallel(t')
=\exp\left[\int_{t'}^0
dt\, s(t)\right] \,, \quad
e_\perp^{d-1}=e_\parallel^{-1} \,.
\label{l10} \end{equation}
Now, the equations (\ref{l5},\ref{l6}) can be rewritten as
\begin{eqnarray} &&
\partial_t\tilde\theta=-i\int d{\bf r}' \,
\chi\big(R(t)\big)\tilde p(t,{\bf r}') \,,
\label{l11} \\ &&
\partial_t\tilde p
=y\delta(t)\delta_\Lambda({\bf r}) \,
\to \,\tilde p=-y \delta_\Lambda({\bf r})
\,, \quad  t<0 \,,
\label{l13} \end{eqnarray}
where we presented an obvious solution for $\tilde p$ satisfying
$\tilde p=0$ at $t>0$. The quantity $R$ in (\ref{l11}) is
\begin{equation}
R^2=e_\parallel^2(x-x')^2+
e_\perp^2({\bf r}_\perp-{\bf r}_\perp')^2 \,.
\label{l12} \end{equation}
Note that
\begin{equation}
\partial_t R=
-s(x\partial_x+x'\partial_x')R
+\frac{s}{d-1}({\bf r}_\perp{\bbox\nabla}_\perp+
{\bf r}'_\perp{\bbox\nabla}'_\perp)R \,.
\label{ly1} \end{equation}
For the considered uniaxial geometry, the relation (\ref{b3}) gives
\begin{equation}
s=-iD\int d{\bf r}\, \tilde p
\left[(d-1)x\partial_x-
{\bf r}_\perp{\bbox\nabla}_\perp\right]\tilde\theta \,.
\label{l14} \end{equation}
Using now (\ref{l11},\ref{l13}) we find
\begin{equation}
\partial_t s=-Dy^2
\int d{\bf r}\,d{\bf r}'\,
\delta_\Lambda({\bf r})\delta_\Lambda({\bf r}')
\left[(d-1)x\partial_x-
{\bf r}_\perp{\bbox\nabla}_\perp\right]\chi(R) \,.
\label{l16} \end{equation}
By virtue of (\ref{ly1}) and the symmetry properties
of the integrand in (\ref{l16}), we obtain
\begin{equation}
s\partial_t s=\frac{(d-1)Dy^2}{2}\partial_t
\int d{\bf r}\,d{\bf r}'\,
\delta_\Lambda({\bf r})\delta_\Lambda({\bf r}')\chi(R) \,.
\label{ly2} \end{equation}
The equation has an obvious first integral which can be
established if to take into account that $s\to 0$ if $t\to-\infty$
[otherwise (\ref{bi3}) is infinite]:
\begin{equation}
s^2=(d-1)Dy^2\int d{\bf r}\,d{\bf r}'\,
\delta_\Lambda({\bf r})\delta_\Lambda({\bf r}')\chi(R) \,.
\label{ly3} \end{equation}

One can demonstrate that the main contribution to ${\cal Z}_\Lambda$
is determined by the saddle point with $s>0$. Then, in accordance with
(\ref{l10}), $e_\parallel$ increases with increasing $|t|$. That means that
the characteristic value of $R$ in (\ref{ly3}) can be estimated as
$R\sim\Lambda^{-1}e_\parallel$. At small $|t|$ where $e_\parallel$ is
not very large, $\chi(R)$ in (\ref{ly3}) can be substituted by
$P_2=\chi(0)$ and we find that $s\simeq s_1$ where
\begin{equation}
s_1=y\sqrt{(d-1)P_2D} \ .
\label{l18} \end{equation}
That leads to $e_\parallel\approx\exp(s_1|t|)$, which is correct if
$R<L$ what means $|t|\lesssim s_1^{-1}\ln(L\Lambda)$. In the opposite limit
$|t|\gg s_1^{-1}\ln(L\Lambda)$, the value of $s$ tends to zero.

The above analysis shows that the main contribution to ${\cal F}_{\rm extr}$
(\ref{bi3}) is associated with the region of integration
$|t|\lesssim s_1^{-1}\ln(L\Lambda)$ and the first term in (\ref{bi3})
can be written as $(1/2)y^2P_2s_1^{-1}\ln(L\Lambda)$. Substituting
(\ref{lf1}) with (\ref{l18}) into
the second term of (\ref{bi3}), we find
\begin{equation}
{\cal F}_{\rm extr}=
\sqrt\frac{P_2y^2}{(d-1)D} \ln(L\Lambda) \,.
\label{hi3} \end{equation}
Note that the expression is in agreement with (\ref{bi8}) since we
considered the case $r_d\Lambda\ll 1$ where the answer  should be
$\kappa$-independent. It is also possible to restore ${\cal Z}_\Lambda(y)$
in the limit $\Lambda\to\infty$ that is for the single-point
object. For this we should recognize that generally
${\cal F}_{\rm extr}$ is a function of the dimensionless parameter
$\Lambda r_d$ and use the property (\ref{bi8}). Then in the limit
$r_d\Lambda\gg 1$ where $\Lambda$-dependence should disappear we obtain
\begin{equation}
{\cal F}_{\rm extr}=
\sqrt\frac{P_2y^2}{(d-1)D}
\left\{\ln(L/r_d) + \frac{1}{4}
\ln\left(\frac{P_2}{D}y^2(d-1)\right)\right\} \,.
\label{hi3'} \end{equation}
Note the nontrivial dependence of this single-point object on $y$.
It is the consequence of the time-dependence of the effective diffusion
cut-off which could be seen at the direct solution with an explicit
account of the diffusion.

Above, we considered the simplest case of the uniaxial strain matrix
$\sigma_{\alpha\beta}$. It is not very difficult to generalize the scheme
for the case where principal axes of $\sigma_{\alpha\beta}$ are fixed
(that is do not depend on time). The answer shows that it is the uniaxial
solution that gives the minimum value of ${\cal F}_{\rm extr}$ and therefore
only this contribution should be taken into account.

As long as we are interesting in the tail of the generating function
${\cal Z}_\Lambda(y)$ at large $y$, the instanton contribution (\ref{hi3})
or (\ref{hi3'}) gives the correct answer.
However, it is not enough to consider that
contribution to obtain the tails of the PDF because the respective tail of
${\cal Z}(y)$ is exponential. Indeed, we shall see below that the tails of
${\cal P}(\theta)$ are determined by the contributions at moderate $y$.
We thus face the problem of finding ${\cal Z}(y)$ at arbitrary $y$.
Fortunately, the tails of ${\cal P}(\theta)$ are also determined by the
instanton contribution yet this instanton is different from the above uniaxial
solution, which represents the situation where stretching occurs along one
marked direction. It is obvious that if the direction slowly varies
in time the value of the effective action will not be essentially
influenced. The role of such soft fluctuations is expected to be
negligible if the characteristic time $s_1^{-1}\ln(L\Lambda)$ of
the stretching is small enough. We thus conclude, taking into account
(\ref{l18}), that the expression (\ref{hi3}) is correct at large $y$.
At moderate $y$, the fluctuations of the stretching direction should
be taken into account, it is the topic of the next subsection. There, we shall
explicitly integrate over soft mode and obtain
different equations for the instanton.

\subsection{Isotropic Instanton}

Here, we are going to take into account the fluctuations of the
stretching direction which were neglected in the preceding
subsection. For that purpose, it is useful to introduce the variable
measuring the stretching rate along the current stretching direction
(the direction of the maximal Lyapunov exponent) determined by the
strain field $\sigma_{\alpha\beta}=\nabla_\beta v_\alpha$. For this aim,
it is useful to perform the transformation of the fields $\theta$, $p$
generalizing (\ref{lf2}) for an arbitrary $\sigma_{\alpha\beta}$
\cite{95CFKLa}.
Namely, let us pass to the fields
\begin{equation}
\tilde\theta(t,{\bf r})
=\theta(t, M_{\alpha\beta}r_\beta)\,, \quad
\tilde p(t,{\bf r})=p(t, M_{\alpha\beta}r_\beta)\,,
\label{gq1} \end{equation}
with $d\times d$ matrix $M_{\alpha\beta}$ controlled by the equation
\begin{equation}
\partial_t\hat M=\hat\sigma\hat M\,, \quad
\hat M(t=0)=\hat 1 \ ,
\label{g1} \end{equation}
with a formal solution
\begin{equation}
\hat M={\rm T}\exp\Big(\int\limits_0^t dt'
\hat\sigma(t')\Big) \ .
\label{g2} \end{equation}
The symbol ${\rm T}$ designates the anti-chronological
ordering for negative $t$.
Note that ${\rm det}\hat M=1$ due to the incompressibility
condition ${\rm tr}\,\hat\sigma=\nabla_\alpha v_\alpha=0$.
Performing the substitution in (\ref{l4}) and passing to the new space
variable $\hat M{\bf r}$ (the Jacobian of the transformation is equal
to unity due to ${\rm det}\hat M=1$), one obtains
\begin{equation}
i{\cal I}=i\int dt\,d{\bf r}\,\tilde p\partial_t\tilde\theta
-\frac{1}{2}\int dt\,d{\bf r}_1\,d{\bf r}_2\,
\tilde p_1\chi(R)\tilde p_2 \,,
\label{g4} \end{equation}
where
\begin{equation}
R_\alpha=M_{\alpha\beta}(r_{1\beta}-r_{2\beta}) \,.
\label{g3} \end{equation}

We see that only ${\bf R}$ is $\sigma$-dependent in (\ref{g4}) and,
moreover, only it's absolute value $R$ enters the effective action.
Just that value is a measure of the stretching irrespective of the directions
of the
current main axes of the matrix $\hat\sigma$.
The statistics of $R$ can be established starting from the PDF
$\exp(-{\cal F})$, see e.g. \cite{94CGK}. The answer is that, for
negative times, $R$ can be written as
\begin{equation}
R(t)=\exp\Big(\int_{t}^0dt'\zeta(t')\Big)
|{\bf r}_1-{\bf r}_2| \,,
\label{g5} \end{equation}
with the random variable $\zeta$ having PDF
$\exp(-{\cal F}_\zeta)$ with
\begin{equation}
{\cal F}_\zeta=\int dt \,
\frac{1}{2D(d-1)}\left(\zeta-
\frac{d(d-1)}{2}D\right)^2 \,.
\label{g6} \end{equation}
The generating functional (\ref{bb6}) is thus rewritten as
\begin{equation}
{\cal Z}_\Lambda
=\int{\cal D}\tilde\theta\,
{\cal D}\tilde p\,{\cal D}\zeta\,
\exp\left(iy\theta_\Lambda+i{\cal I}-{\cal F}_\zeta\right) \,,
\label{g7} \end{equation}
where $\theta_\Lambda$ is defined by (\ref{bb7}).

We have performed the exact transformation of the statistical weight.
Let us stress that the situation described by (\ref{g7})
is isotropic from the beginning whereas the solution found in the
preceding subsection was anisotropic.
Now, we formulate extremum conditions describing a
saddle point for the argument of the exponent in (\ref{g7}):
\begin{eqnarray} &&
\partial_t\tilde p=y\delta(t)\delta_\lambda({\bf r}) \,,
\label{g8} \\ &&
\partial_t\tilde\theta(t,{\bf r}_1)=
-i\int d{\bf r}_2\, \chi[R(t)]\tilde p(t,{\bf r}_2) \,,
\label{g9} \\ &&
\zeta(t')=\frac{d(d-1)}{2}D
-\frac{d-1}{2}D\int\limits_{-\infty}^{t'}
dt\,d{\bf r}_1\,d{\bf r}_2\,
\tilde p_1\tilde p_2\frac{\partial\chi}{\partial R}R \,.
\label{g10} \end{eqnarray}
The equation (\ref{g8}) has the same form as (\ref{l13}) and has
consequently the same solution $\tilde p=-y\delta_\Lambda({\bf r})$.
It follows from (\ref{g7},\ref{g8}) that in the saddle-point approximation
${\cal Z}_\Lambda\propto\exp(-{\cal F}_{\rm extr})$ where
\begin{equation}
{\cal F}_{\rm extr}=
\frac{1}{2}\int dt\,d{\bf r}_1\, d{\bf r}_2\,
\tilde p_1 \tilde p_2 \chi(R)
+\frac{1}{2D(d-1)}\int dt\,
\left(\zeta-\frac{(d-1)d}{2}D\right)^2 \,.
\label{g15} \end{equation}

It follows from (\ref{g5}) that $\partial_t R=-\zeta R$. Using that,
we can find the first integral of the equation (\ref{g10}):
\begin{equation}
\zeta^2=\frac{d^2(d-1)^2}{4}D^2
+(d-1)Dy^2\int d{\bf r}_1\, d{\bf r}_2\,
\delta_\Lambda({\bf r}_1)\delta_\Lambda({\bf r}_2)\chi(R) \,.
\label{g12} \end{equation}
The constant here is established using the property $\zeta\to(d-1)dD/2$
at $t\to-\infty$ following from $\theta\to0$ at $t\to-\infty$
[(\ref{g15}) is infinite otherwise]. The characteristic value of $R$ in
the right-hand side of (\ref{g12}) can be estimated as
\begin{equation}
R(t')\sim\Lambda^{-1}\exp\int\limits_{t'}^0dt\,\zeta(t) \,.
\label{g13} \end{equation}
If $R\ll L$ then the integral in the right-hand side of (\ref{g12}) is
approximately equal to $P_2$, if $R\gg L$ then the integral is negligible.
That means that there are two different time intervals. At large $|t|$, it is
$\zeta\simeq (d-1)dD/2$ and at small $|t|$ it is $\zeta\simeq\zeta_1$ where
\begin{equation}
\zeta_1^2=\frac{(d-1)^2d^2}{4}D^2
+(d-1)Dy^2P_2 \,.
\label{g14} \end{equation}
The boundary between the regions is at $|t|\sim \zeta_1^{-1}\ln(L\Lambda)$.
The main contribution to ${\cal F}_{\rm extr}$ (\ref{g15}) is associated
with the region $|t|<t_1=\zeta_1^{-1}\ln(L\Lambda)$. The first term in
(\ref{g15}) can be substituted by $y^2P_2t_1/2$ and the second
one can be substituted by
$\big(2D(d-1)\big)^{-1}\big(\zeta_1-(d-1)dD/2\big)^2t_1$.
Using (\ref{g14}) we find
\begin{equation}
{\cal F}_{\rm extr}=\left(\sqrt{
\frac{d^2}{4}+\frac{P_2y^2}{(d-1)D}}
-\frac{d}{2}\right)\ln(L\Lambda) \,.
\label{g16} \end{equation}
Comparing the expression (\ref{g16}) with (\ref{hi3}), we conclude that
fluctuations of the stretching direction can be neglected if
$y^2\gg Dd^3P_2^{-1}$. Let us stress that at $y^2\sim Dd^3P_2^{-1}$
the value of (\ref{g16}) is much larger than unity. That means that
violation of (\ref{hi3}) is not associated with destructing
saddle-point regime, it is rather related to an incorrect
calculation of soft fluctuations in the saddle-point regime. Note also
that the role of fluctuations increases with increasing the space
dimensionality
$d$.

\subsection{Probability Distribution Function}

The scheme proposed in the preceding subsections can be applied
also to calculating PDF ${\cal P}_\Lambda(\vartheta)$ of the quantity
$\theta_\Lambda$ (\ref{bb7}). Let us start from the average
\begin{equation}
\langle\theta_\Lambda^{2n}\rangle=
\int{\cal D}\theta\,{\cal D}p\,{\cal D}\sigma\,
\exp(i{\cal I}-{\cal F}_\sigma +2n\ln\theta_\Lambda) \,.
\label{g29} \end{equation}
Thus we see that the saddle-point contribution to
$\langle\theta_\Lambda^{2n}\rangle$ is determined by extrema of
$i{\cal I}-{\cal F}_\sigma +2n\ln\theta_\Lambda$ which coincide
with (\ref{b1},\ref{b2},\ref{b3}) if to substitute
\begin{equation}
y\to -\frac{2ni}{\theta_\Lambda} \,.
\label{g30} \end{equation}
Then an attempt to find the analog of the uniaxial instanton
fails. The formal reason for this is in additional $i$ in (\ref{g30}).
The physical reason is that the uniaxial instanton is an adequate
object for the statistics of fast processes whereas
$\langle\theta_\Lambda^{2n}\rangle$ is determined by slow processes.

To find a solution, we should pass to the isotropic instanton.
That means that we should perform the same transformation of the fields
as in the preceding subsection what leads to the saddle-point equations
(\ref{g8},\ref{g9},\ref{g10}) with (\ref{g30}). The equations has a solution
of the same type as considered above with
\begin{equation}
\zeta_1^2 =\frac{d^2(d-1)^2}{4}D^2
-(d-1)DP_2 \frac{4n^2}{\theta_\Lambda^2}\,.
\label{g41} \end{equation}
The value of $\theta_\Lambda$ in (\ref{g41}) is the parameter which can
be found from the equation analogous to (\ref{g9}) which now reads
\begin{equation}
\partial_t\tilde\theta
=\frac{2n}{\theta_\Lambda}
\int d{\bf r}_2\chi(R)\delta_\Lambda({\bf r}_2) \,.
\label{g31} \end{equation}
As previously, the integral in the right-hand side of (\ref{g31})
for ${\bf r}_1=0$ is equal to $P_2$ if $|t|\lesssim\ln(L\Lambda)/\zeta_1$
and is negligible otherwise. We thus come to the conclusion that
\begin{equation}
\theta_\Lambda^2\simeq \theta^2(t=0,{\bf r}=0)
\simeq 2n P_2\frac{\ln(L\Lambda)}{\zeta_1} \,.
\label{g32} \end{equation}
Substituting the relation into (\ref{g41}) we find the equation
on $\zeta_1$ leading to
\begin{equation}
\zeta_1=(d-1)D\left\{-\frac{n}{\ln(L\Lambda)}
+\sqrt{\frac{n^2}{\ln^2(L\Lambda)}
+\frac{d^2}{4}}\right\} \,.
\label{g33} \end{equation}
We see that $\zeta_1$ decreases with increasing $n$ and consequently
the characteristic time $\ln(L\Lambda)/\zeta_1$ increases with
increasing $n$. Substituting now (\ref{g33}) into (\ref{g32})
one obtains
\begin{equation}
\theta_{\Lambda n}^2
=\frac{8nP_2\ln(L\Lambda)}{d^2D(d-1)}
\left\{\frac{n}{\ln(L\Lambda)}
+\sqrt{\frac{n^2}{\ln^2(L\Lambda)}
+\frac{d^2}{4}}\right\} \,.
\label{g34} \end{equation}

It is not very difficult to recognize that the main
contribution to the saddle-point value of
$i{\cal I}-{\cal F}_\zeta +2n\ln\theta_\Lambda$
is determined by the last term. That means that
\begin{equation}
\langle\theta_\Lambda^{2n}\rangle\propto
\exp(-{\cal F}_{\rm extr})\propto
\theta_{\Lambda n}^{2n} \,,
\label{g35} \end{equation}
with $\theta_{\Lambda n}$ from (\ref{g34}).

The same result can be deduced by the alternative method.
Namely, starting from (\ref{g16}) we can calculate the tail of the PDF
${\cal P}_\Lambda(\vartheta)$ for the quantity $\theta_\Lambda$ (\ref{bb7}).
The function ${\cal P}_\Lambda(\vartheta)$ is the Fourier transform
of ${\cal Z}_\Lambda(y)$:
\begin{equation}
{\cal P}_\Lambda(\vartheta)=\int dy\,
\exp(-iy\vartheta){\cal Z}_\Lambda(y)
\propto \int dy\, \exp(-iy\vartheta-{\cal F}_{\rm extr}) \,.
\label{g17} \end{equation}
Substituting here (\ref{g15}) and calculating the integral over
$y$ by the saddle-point method \cite{94CGK} we find
\begin{equation}
{\cal P}_\Lambda(\vartheta)\propto
\exp\left\{\frac{d}{2}\ln(L\Lambda)\left(
1-\sqrt{1+\frac{d-1}{P_2}D
\frac{\vartheta^2}{\ln^2(L\Lambda)}}
\right)\right\} \,,
\label{g18} \end{equation}
what is in agreement with \cite{94SS,94CGK,95CFKLa}.
Formally, the expression (\ref{g18}) is valid at $\vartheta\to\infty$ but
really it covers the whole region of $\vartheta$
because the PDF is Gaussian at small $\vartheta$ \cite{95CFKLa}. The distant
tails
of the PDF are exponential as has been established first by Shraiman and
Siggia \cite{94SS}.
Note that the value of the Lyapunov exponent $\zeta_1$ corresponding
to the saddle point in (\ref{g17}) is
\begin{equation}
\zeta_{\rm extr}=
\frac{d}{2}(d-1)D
\left(1+\frac{d-1}{P_2}D
\frac{\vartheta^2}{\ln^2(L\Lambda)}\right)^{-1/2}\,.
\label{g19} \end{equation}
We see that the value decreases with increasing $\vartheta$
whereas the value $\zeta_1$ (\ref{g14}) increases with increasing $y$.
Note also that the value of $y$ corresponding to the extremum point is
\begin{equation}
y_{\rm extr}^2=
-\frac{(d-1)d^2D}{4P_2}
\frac{\vartheta^2}{\vartheta^2+P_2/((d-1)D)} \,.
\label{g20} \end{equation}
That means that $|y_{\rm extr}^2|<d^3D/P_2$ and consequently the
extremum point lies beyond the applicability region of the approximation
(\ref{hi3}). This is the reason why (\ref{hi3}) does not admit to restore
${\cal P}_\Lambda(\vartheta)$.

Now, we can calculate $\langle\theta_\Lambda^{2n}\rangle$ starting
from the definition
\begin{equation}
\langle\theta_\Lambda^{2n}\rangle=
\int d\vartheta \, \vartheta^{2n}
{\cal P}_\Lambda(\vartheta) \,.
\label{g36} \end{equation}
This integral can be calculated using again the saddle-point
method. The result coincides, of course, with (\ref{g35}). We thus
conclude that (\ref{g15}) or (\ref{g18}) cover both cases of slow
and of fast processes. That means that the account of fluctuations
in the direction of stretching (performed in the preceding subsection)
give us the tool for finding tails of both PDF and of the generating
functional.

\subsection{Discussion}

We considered the statistics of the passive scalar advected by the
random velocity field in the framework of the instanton formalism.
The consideration was very instructive since it revealed some
nontrivial peculiarities of the formalism. First of all, we see that
the direct solution of the saddle-point equations gives us the answer
which satisfactory describes the tail of the generating functional
${\cal Z}(\lambda)$ but cannot serve to restore the tail of PDF
${\cal P}(\vartheta)$. The physical reason for this lies in difference of
processes forming the tails: The tail by ${\cal Z}$ is related to
the fast processes with the characteristic time decreasing as $\lambda$
increases while the tail by ${\cal P}$ is related to slow
processes with characteristic time increasing as $\vartheta$ increases.
The conclusion can be directly extracted from (\ref{l18})
and (\ref{g19}). For the slow processes, the fluctuations of the stretching
direction are relevant which do not destroy the saddle-point (instanton)
regime but renormalize the naive answer. For that particular problem, the
fluctuations can be explicitly taken into account after the special
transformations of the fields. Although the trick cannot be widely
generalized it shows the direction of improving naive answers. In the
general case, we expect that the direct solution of the saddle-point
equations will produce nonsymmetric instantons with a degeneracy
parameter (like the direction of the marked axis in the considered case).
Then, there exists the ``Goldstone'' mode related to slightly nonhomogeneous
variations of the parameter. Such mode is soft since it only weakly disturbs
the naive instantons. Therefore, the fluctuations related to the soft mode
are relevant and should be explicitly taken into account.

\section{The simplest instanton of an incompressible velocity field}
\label{sec:nse}

Here, we describe the first step in considering a much more
complicated problem of finding the tails
of the PDF for velocity field in three-dimensional incompressible
turbulence. We consider two-point statistics and show that an instanton
with a linear spatial profile naturally appears as a basic flow.

The effective action (\ref{i4}) for the Navier-Stokes equation
can be written as follows
\begin{eqnarray} &&
{\cal I}=\int dt\, d{\bf r}\,
(p_\alpha\partial_t v_\alpha +
p_\alpha v_\beta\nabla_\beta v_\alpha-\nu p_\alpha\nabla^2 v_\alpha
+p_\alpha \nabla_\alpha P +Q \nabla_\alpha v_\alpha)
\nonumber \\ &&
+{i\over2}\int dt\, dt'\, d{\bf r}\, d{\bf r}'\,
\Xi(t-t',{\bf r}-{\bf r}') p_\alpha p'_\alpha \,,
\label{j4} \end{eqnarray}
The
additional independent fields $P$ and $Q$ play the role of
Lagrange multipliers enforcing the incompressibility conditions
$\nabla_\alpha v_\alpha=0$ and the analogous condition
$\nabla_\alpha p_\alpha=0$ for the response field $p_\alpha$.
The field $P$ has the meaning of pressure (divided by the mass density $\rho$).
The origin of the terms with the fields $P$, $Q$ in the effective action is
related to the continuity equation
$\partial_t\rho+\nabla_\alpha(\rho v_\alpha)=0$,
which should be incorporated
into the effective action, $Q$ is just the auxiliary (response) field
corresponding to the equation. At the condition that all velocities are
much smaller than the sound velocity, it is possible to neglect the
time derivative in $\partial_t\rho+\nabla_\alpha(\rho v_\alpha)=0$
and variations of the mass density what leads to the term
$Q\nabla_\alpha v_\alpha$ in (\ref{j4}). While variations of the mass
density can be neglected variations in the pressure are relevant.
Therefore, it is natural to pass from the integration over the mass
density to the integration over the pressure as it is implied in (\ref{j4}).

We are going to examine the generating functional for the velocity
\begin{eqnarray} &&
{\cal Z}({\bbox\lambda})\equiv
\left\langle \exp\left(i\int dt\,d{\bf r}\,
{\bbox\lambda}{\bf v}\right)\right\rangle
\nonumber \\ &&
=\int {\cal D}p\,{\cal D}v\,{\cal D}P\,{\cal D}Q\,
\exp\left(i{\cal I}+i\int dt\,d{\bf r}\,
{\bbox\lambda}{\bf v}\right) \,.
\label{j5} \end{eqnarray}
The extremum conditions for the argument of the exponent in (\ref{j5})
determining the Navier-Stokes instanton read
\begin{eqnarray}  &&
\partial_t{v}_\alpha({\bf r})
+ v_\beta(t,{\bf r}) \nabla_\beta v_\alpha(t,{\bf r})
-\nu \nabla^2 v_\alpha(t,{\bf r})
+\nabla_\alpha P(t,{\bf r})=
-i\int dt'\,\int\frac{d^dk}{(2\pi)^d}\exp(i{\bf kr})
\,\Xi(t-t',{\bf k})\, p_\alpha(t',{\bf k}) \,,
\label{eu1} \\  &&
\partial_t{p}_\alpha(t,{\bf r})
-p_\beta(t,{\bf r}) \nabla_\alpha v_\beta(t,{\bf r})
+ v_\beta(t,{\bf r}) \nabla_\beta p_\alpha(t,{\bf r})
+ \nu \nabla^2 p_\alpha(t, {\bf r})
+\nabla_\alpha Q(t,{\bf r})
=\lambda_\alpha(t,{\bf r})\,,
\label{eu2} \end{eqnarray}
where $\Xi({\bf k})$ and $p_\alpha({\bf k})$ are Fourier transforms
of $\Xi({\bf r})$ and $p_\alpha({\bf r})$ respectively.
In (\ref{eu1},\ref{eu2}) the conditions $\nabla_\alpha v_\alpha=0$,
$\nabla_\alpha p_\alpha=0$ are also implied which originate from varying
over the fields $P$ and $Q$. Then the values of the fields $P$ and $Q$ can
also be found from the conditions. It gives the relations
\begin{eqnarray} &&
\nabla^2 P= -\nabla_\alpha
(v_\beta \nabla_\beta v_\alpha) \,.
\label{j6} \\ &&
\nabla^2 Q=\nabla_\alpha(p_\beta \nabla_\alpha v_\beta
- v_\beta \nabla_\beta p_\alpha) \,.
\label{jj6} \end{eqnarray}
Note that similar idea of the instanton formalism has been discussed also by
Giles \cite{Gil} yet his approach is based on uncontrollable approximations.

In the following we consider the simultaneous correlation functions of
the velocity differences $\langle [{\bf v}(0,{\bbox\rho/2})-
{\bf v}(0,-{\bbox\rho}/2)]^{2n}\rangle$ where ${\bbox\rho}$ is
the separation between the points. The functional generating such
functions is extracted from ${\cal Z}(\lambda)$ if one gets
\begin{equation}
\lambda_\alpha=y n_\alpha \delta(t)
[\delta({\bf r}-{\bbox\rho}/2)-
\delta({\bf r}+{\bbox\rho}/2)] \,,
\label{j7} \end{equation}
where ${\bf n}$ is a unit vector. As was explained in Introduction
the presence of such term in right-hand side of (\ref{eu2}) means that
we should solve the problem at negative times $t$ with the final condition
\begin{equation}
p_\alpha(0,{\bf r})=-y
(\delta_{\alpha\beta}-\nabla_\alpha\nabla_\beta \nabla^{-2})
n_\beta[\delta({\bf r}-{\bbox\rho}/2)-
\delta({\bf r}+{\bbox\rho}/2)] \,.
\label{j8} \end{equation}

We assume that the pumping correlation function $\Xi$
is delta-correlated in time: $\Xi(t,{\bf r})=\delta(t)\chi({\bf r})$.
Then the system of equations (\ref{eu1}-\ref{jj6}) is invariant under
the transformation analogous to (\ref{bi8})
\begin{equation}
t\to X^{-1}t\,, \quad {\bf v}\to X{\bf v}\,, \quad
P\to X^{2}P\,, \quad Q\to X^{3}Q\,, \quad
\nu\to X\nu\,, \quad \lambda\to X\lambda \,, \quad
{\bf p}\to X^{3}{\bf p} \,,
\label{jq8} \end{equation}
where $X$ is an arbitrary factor. For the function (\ref{j8}) the
transformation (\ref{jq8}) means $y\to X^2 y$. The extremum value
${\cal F}_{\rm extr}$ of the argument of the exponent in (\ref{j5})
transforms as ${\cal F}_{\rm extr}\to X^3 {\cal F}_{\rm extr}$ at
(\ref{jq8}). That leads to the conclusion that
\begin{equation}
{\cal Z}(y)\propto\exp(-{\cal F}_{\rm extr})\, \quad
{\cal F}_{\rm extr}=y^{3/2}f(y/\nu^2) \,,
\label{jq9} \end{equation}
with the function $f$ to be determined. We expect that in the limit
$y\to\infty$ a $\nu$-dependence in the function disappears. Then
we conclude ${\cal F}_{\rm extr}\propto y^{3/2}$.

The characteristic wave vector $k_0$ in the correlation function
$\chi({\bf k})$ of the pumping force is of the order of the
inverse pumping length $L$.
Then examining the region ${\bf r}\ll L$ one can expand the exponent
$\exp(i{\bf kr})$ in (\ref{eu1}) into the series over ${\bf kr}$.
The first term of the expansion produces the zero contribution
to the right-hand side of (\ref{eu1}) because of the structure
of the field $p$ determined by the condition (\ref{j8}): The
condition means that at $t=0$ ${\bf p}({\bf r})=-{\bf p}(-{\bf r})$,
the property is reproduced by the equations, so that
${\bf p}({\bf k})=-{\bf p}(-{\bf k})$ at any time $t$. Thus the
leading term of the expansion of the right-hand side is linear in
${\bf r}$. That means that the equation (\ref{eu1}) admits as a
solution in the region $|{\bf r}|\ll L$ a linear profile
\begin{equation}
v_\alpha=\sigma_{\alpha\beta}(t)r_\beta \,,
\quad \sigma_{\alpha\alpha}=0 \,.
\label{j9} \end{equation}
Then we obtain from (\ref{eu1})
\begin{equation}
\partial_t\sigma_{\alpha\gamma}
+\sigma_{\alpha\beta}\sigma_{\beta\gamma}
-\frac{1}{d}\delta_{\alpha\gamma}
(\sigma_{\mu\nu}\sigma_{\nu\mu})
=\int \frac{d^dk}{(2\pi)^d}
k_\gamma p_\alpha({\bf k})\chi({\bf k}) \,.
\label{j10} \end{equation}
Here we substituted the expression for the pressure
\begin{equation}
P=-\frac{1}{d}(\sigma_{\mu\nu}\sigma_{\nu\mu})r^2 \,,
\label{j11} \end{equation}
which provides for the condition $\nabla_\alpha v_\alpha=0$. Note that $P$
is defined up to a harmonic function, the expression (\ref{j11}) is
chosen because of its isotropy.

For the linear velocity profile the equation (\ref{eu2}) can be rewritten
in Fourier representation as
\begin{equation}
\partial_t p_\alpha-\sigma_{\beta\alpha}p_\beta
-\sigma_{\beta\gamma}k_\beta\frac{\partial}{\partial k_\gamma}p_\alpha
-\nu k^2 p_\alpha +i k_\alpha Q= 0
\label{j12} \end{equation}
which should be solved with the condition following from (\ref{j8}):
\begin{equation}
p_\alpha(t=0,{\bf k})=2iy \left(\delta_{\alpha\beta}
-\frac{k_\alpha k_\beta}{k^2}\right)
n_\beta\sin({\bf k}{\bbox\rho}/2) \,.
\label{j13} \end{equation}
The characteristic wave vector ${\bf k}$ in (\ref{j10}) is of the order
of $L^{-1}$. Thus we can expand $\sin({\bf k}{\bbox\rho}/2)$ in
${\bf k}{\bbox\rho}$ and keep only the first nonvanishing term of the expansion
$\propto{\bf k}{\bbox\rho}$. As was discussed in
the Introduction, the response field ${\bf p}({\bf r},t)$
propagates backwards in time, starting with the initial value (\ref{j13})
at $t=0$. We shall see that for a long time (determined
by a small viscosity) the field ${\bf p}(t,{\bf k})$ at $k\sim L^{-1}$
has the same structure $\propto{\bf k}{\bbox\rho}$.

There is a general family of the flows with linear profiles --
see Sect.\ref{sec:fam} below. We start by considering the simplest case.
We assume below that the point separation $\bbox\rho$ is directed along
the same vector $n_\alpha$ as the measured velocity components:
$\rho_\alpha = n_\alpha \rho$. Then the problem possesses the
axial symmetry which allows us to look for the following uniaxial
strain matrix
\begin{equation}
\sigma_{\alpha\beta} = s \left({\delta_{\alpha\beta}
-d n_\alpha n_\beta}\right) \,.
\label{j18} \end{equation}

The same symmetry admits the anzatz
\begin{equation}
p_\alpha(t,{\bf k})=\left(\delta_{\alpha\beta}
-\frac{k_\alpha k_\beta}{k^2}\right)
iy n_\beta\phi(t,z){\bf kn}\rho \,,
\label{j14} \end{equation}
correct in the limit of small $k$. In (\ref{j14}) $z={\bf kn}/k$ and
the function $\phi(t,z)$ to be found has the initial (final) value
$\phi(t=0,z)=1$. Substituting (\ref{j13},\ref{j14}) into (\ref{j12})
we find $Q=0$ and
\begin{equation}
s^{-1}(\partial_t-\nu k^2)\phi
+dz(1-z^2)\partial_z\phi
+2[(d-1)-dz^2]\phi=0 \,.
\label{j20} \end{equation}

Substituting the expression (\ref{j14}) into the
right-hand side of (\ref{j10}) we find
\begin{equation}
\int \frac{d^dk}{(2\pi)^d}
k_\gamma p_\alpha({\bf k})\chi({\bf k})=
G(dn_\alpha n_\gamma-\delta_{\alpha\gamma}) \,,
\label{j15} \end{equation}
where
\begin{equation}
G=iy\rho C \int\limits_{-1}^1 dz\,
z^2(1-z^2)^{(d-1)/2}\phi(t,z) \,,
\label{j16} \end{equation}
and
\begin{equation}
C=\frac{S_{d-1}}{(2\pi)^d(d-1)}
\int\limits_0^\infty dk\, k^{d+1}\chi(k) \,.
\label{j17} \end{equation}
Here $S_d$ is the area of the unit sphere in $d$-dimensional space
$S_d=2\pi^{d/2}/\Gamma(d/2)$. The constant $C$ can be estimated as
$C\sim{\cal E}/L^2$ where ${\cal E}
=\langle{v_\alpha\partial_t{v}_\alpha}\rangle$
is the energy dissipation rate. Substituting now (\ref{j18},\ref{j15})
into (\ref{j10}) we find
\begin{equation}
\partial_t{s} = (d-2) s^2  - G.
\label{j19} \end{equation}

Our next problem is to find $G$ as a functional of s, to close this set
of equations.  We have to solve the equation (\ref{j20}) for $\phi$.
Here, viscous and inviscid cases are slightly different. We start
from considering an inviscid Euler equation, then we will account for
the viscosity.

\subsection{Instanton of the Euler equation}

Neglecting viscosity in (\ref{j20}) we get a general solution
\begin{equation}
{\phi}=h^{2}z^{-2}F\left({z^2h^{-2d}\over 1-z^2}\right)\ ,\quad
h(t)=\exp\biggl(
\int_0^t s(t')dt'\biggr)\ .
\end{equation}
The initial condition $\phi(0,z)=1$ fixes the function $F$:
\begin{equation}
\phi(z) = \frac{h^{2- 2 d}}{{1-z^2 + z^2 h^{-2d}}} \,.
\end{equation}
We get the system of equations
\begin{equation}
\dot{s} = (d-2)s^2 - G(h),\quad
\dot{h} = s h\ .
\label{dots} \end{equation}
This system  for the variable $q = h^{2-d}$ becomes the usual potential
problem $\ddot{q} = -U'(q)$ with the potential
\begin{eqnarray}  & &
U(q) = -(d-2) \int d q \, q G\left({q^{1\over{2-d}}}\right)\nonumber\ .
\end{eqnarray}
For $d=3$,
\begin{eqnarray}
G(h)={iy\rho Ch^2\over h^{2d}-1}
\left[{2\over 1-h^{-2d}}-{1\over3}-{\ln\bigl(2h^{2d}-1
\bigr)\over h^{2d}-1}\right]\nonumber\end{eqnarray}

The relevant solution, which vanishes at $t = -\infty$, corresponds to the zero
energy in this potential [$s(t)\propto -1/t$ as $t\rightarrow-\infty$].
Therefore, $\dot{q}^2 = 2U(0) - 2U(q)$ and
\begin{equation}
{\dot{q}}_{t=0} = \sqrt{2[U(0) -U(1)]} = C_1 \sqrt{{\cal E} y  \rho}/L.
\end{equation}
Then, the strain at the moment $t=0$ becomes
$\sigma_{\alpha\beta} = {\dot{q}}({\delta_{\alpha\beta}- d
n_\alpha n_\beta})/q({2-d})$. In accordance with (\ref{ij})
the logarithmic derivative of ${\cal Z}$-functional is related
to the average initial value  of the velocity difference
${{\cal Z}'( y )}/{{\cal Z}(y)} = n_\alpha \langle{v_\alpha(\rho,0)
- v_\alpha(-\rho,0)}\rangle$.
In the leading WKB approximation at large $ y $ this average can be
replaced by the contribution from the instanton solution:
\begin{equation}
(\ln {\cal Z})'(y) = 2 \rho n_\alpha n_ \beta \sigma_{\alpha\beta} = \frac{2
\dot{q}(d-1)}{q(d-2)} = C_2  \sqrt{{\cal E} y \rho^3L^{-2}}.
\end{equation}
Finally, we obtain the surprisingly simple result
\begin{equation}
{\cal Z}(y) \propto\exp\left[C_3\sqrt{{\cal E}(y\rho)^{3}L^{-2}}
\right].\label{ass}
\end{equation}
with the dimensionless constants $C_1,C_2$ and $C_3$ to be calculated.
This result is in agreement with the general form (\ref{jq9}),
it contains also $\rho$-dependence.

\subsection{Account of viscosity}

When the viscous terms are kept,
the solution is modified as follows.
With the same assumption  $k_0\rho\simeq\rho/L\ll1$,
we can still look for the uniform strain solution. The
viscosity will drop from the velocity equation, but not from the
response field equation (\ref{j12}).
The extra term $\nu k^2p$ can be compensated by extra time dependent
exponential
\begin{equation}
p_\alpha(t,{\bf k})=\left(\delta_{\alpha\beta}
-\frac{k_\alpha k_\beta}{k^2}\right)
iy n_\beta\phi(t,z){\bf kn}\rho \exp\left[{ \nu R(t,z)
k^2}\right]\ .
\end{equation}
The balance of $k^0$ and $k^1$ terms in the equation is the same as
before. The balance of $k^2$ terms gives the equation
\begin{equation}
\dot{R} =1+s\hat LR= 1 -dsz(1-z^2){\partial R\over \partial z}+2s(1-dz^2)R
\end{equation}
with the boundary condition $R(0)=0$.
The substitution $R(z,t)=A(t)+B(t)z^2$
reduces the PDE to two ODEs
\begin{eqnarray}\dot A&=&1+2As\nonumber\\
\dot B&=&2s(B-Bd-Ad)\ .
\nonumber\end{eqnarray}
The solution is expressed via $s(t)$; at  $t\rightarrow-\infty$
it grows linearly: $R\approx t$. The influence of the viscosity
on our solution is weak, it smears the peaks at $p$ and
manifests itself when $\nu R\simeq L^2$ i.e. at $t\simeq L^2/\nu$.
That time should be much larger than the time of instanton formation
$\sqrt{L^2/{\cal E}y\rho}$. Our asymptotic expression (\ref{ass}) is
insensitive
to viscosity if $\nu\ll L\sqrt{{\cal E}y\rho}$.

\subsection{Instanton family}
\label{sec:fam}

Considering more general strain does not change basic conclusions of
this section. Let us describe, for instance, a general three-dimensional
symmetric flow of the type considered in \cite{Ng}. In the cylindrical
coordinates with $z$-axis along ${\bbox\rho}$, the velocity vector field
at $r\ll L$ is given by
\begin{equation}{\bf u}=(u_r,u_\theta,u_z)=(-\sigma r/2,\omega r/2,\sigma z)\ .
\label{sym}\end{equation}
Here vorticity has only $z$-component $\omega(t)$ which is a function
of time as well as strain $\sigma(t)$. The pressure is now of the form
$$P=-gr^2-e[r^2\sin\theta\cos\theta+rz(\sin\theta+\cos\theta)]\ .$$
Particular details of the solution depend on the relation between $g$ and $e$.
The diagonal
elements (proportional to $g$) are determined locally from Poisson equation
$\Delta P=-div({\bf u}\cdot\nabla{\bf u})$. Note that
the off-diagonal pressure elements are generally
determined by the global structure of the flow. In our case, the value of $e$
is determined by the distant
asymptotics $u\rightarrow0,P\rightarrow$const at $r\rightarrow\infty$
and matching conditions at $r\simeq L$, which depends on the particular choice
of
the pumping $\chi$. The global description of the flows for the
whole instanton family is still ahead of us. As far as the functional
dependence
of the respective ${\cal Z}( y ,{\bbox\rho})$ is concerned,
it is the same for the whole
family and does not depend of the large-scale behavior of the pumping.
Considering, for instance, the case $g=0$ [opposite to the diagonal case
(\ref{j11}) considered before],
we get $\omega=\sqrt3\,s$ and the system of equations
similar to (\ref{dots})
$$\dot{s} = s^2 - G'(h),\quad \dot{h} = s h$$
with another yet qualitatively similar function $G'$. For the variable $q(t)$,
related to $s(t)$ by $\dot q=-sq$, the Newton equation appears with a potential
energy $U$ that allow for a single solution (zero energy separatrix) vanishing
as $s(t)\propto 1/t$ at $t\rightarrow-\infty$.
The basic result
$\ln {\cal Z}( y,\rho ) \propto( y  \rho)^{{3}/{2}}\sqrt{\cal E} /L$
is valid for the whole family in agreement with (\ref{jq9}).

\subsection{Discussion}

The particular instanton found has the scaling
\begin{equation}\delta u(\rho)=u(\rho/2)-u(-\rho/2)\propto \rho\ ,\label{sca}
\end{equation}
it would
give the asymptotics of the right tail of the PDF ${\cal P}(\delta
u,\rho)\propto
\exp[- (\delta u/\rho)^3]$. It is unclear at the moment if there are flows
where
such an asymptotics takes place; most probably, this simplest instanton does
not
realize the main extremum of the action.
Note that the similar instanton
with the linear profile is found for the
Burgers problem\cite{GurarieMigdal95} where it indeed determines the right tail
($\delta u>0$) of velocity PDF due to sawtooth waves.
The general analysis of the whole family of
instanton solutions for the two-point velocity statistics at the
framework of the Navier-Stokes equation
will be published elsewhere. Also, the crucial problem of
the contribution to the action from the fluctuations against the
instanton background will be considered. It is clear that, in the straining
flow of the instanton, any
vorticity perturbation produces a spiral with the accumulation point
at the velocity null. The scaling of the perturbation contribution
is different from (\ref{sca}); for instance, it will
give Kolmogorov's $5/3$-law for the
pair correlation function as in the Lundgren example \cite{Lun}.
The analysis of the instanton fluctuations
will be the subject of further publications. Note that the instanton formalism
provides a natural (and long-expected) tool for incorporating numerous
results on particular solutions of the Navier-Stokes equations into the
statistical theory of turbulence.

\acknowledgements
We are grateful to E. Balkovsky and M. Chertkov for useful discussions.
This work was partially supported by the National Science Foundation
under contract PHYS-90-21984 (A.M.), by the Minerva Center for Nonlinear
Physics
(V.L. and I.K.) and by the Rashi Foundation (G.F.).

\end{document}